\documentclass[journal,oneside]{IEEEtran}
\usepackage{cite}

\ifCLASSINFOpdf
\else
   \usepackage[dvips]{graphicx}
\fi
\usepackage[cmex10]{amsmath}
\usepackage{amssymb,amsthm}
\usepackage{balance}

\newtheorem{definition}{Definition} 
\newtheorem{theorem}{Theorem} 
\newtheorem{remark}{Remark}

\begin{document}
%
\title{Accelerating Iterative Detection for Spatially Coupled Systems by Collaborative Training}
%
%
%

\author{Keigo~Takeuchi,~\IEEEmembership{Member,~IEEE}
\thanks{Manuscript received December 10, 2012. 
The associate editor coordinating the review of this letter 
and approving it for publication was J.\ Choi.
}
\thanks{K.~Takeuchi is with the Department of Communication Engineering and 
Informatics, the University of Electro-Communications, Tokyo 182-8585, Japan 
(e-mail: ktakeuchi@uec.ac.jp).}
\thanks{Digital Object Identifier ***}
}

%
%

\markboth{IEEE communications letters,~Vol.~, No.~, 2013}%
{Takeuchi:Accelerating Iterative Detection for Spatially Coupled Systems by Collaborative Training}
%

\IEEEpubid{0000--0000/00\$00.00~\copyright~2013 IEEE}


\maketitle

\begin{abstract}
This letter proposes a novel method for accelerating iterative detection 
for spatially coupled (SC) systems. An SC system is constructed by 
one-dimensional coupling of many subsystems, which are classified into 
training and propagation parts. An irregular structure is introduced into 
the subsystems in the training part so that information in that part can 
be detected successfully. The obtained reliable information may spread over 
the whole system via the subsystems in the propagation part. In order to allow 
the subsystems in the training part to collaborate, {\em shortcuts} between 
them are created to accelerate iterative detection for that part. As an example 
of SC systems, SC code-division multiple-access (CDMA) systems are 
considered. Density Evolution for the SC CDMA systems shows that the proposed 
method can provide a significant reduction in the number of iterations 
for highly loaded systems. 
\end{abstract}


%

\section{Introduction}
\IEEEPARstart{S}{patial} coupling has been proved to improve the 
belief-propagation (BP) performance of conventional low-density parity-check 
(LDPC) codes up to the corresponding maximum-a-posteriori (MAP) 
performance~\cite{Kudekar11}. This phenomenon has been observed in many other 
problems, such as code-division multiple-access (CDMA) 
systems~\cite{Takeuchi111,Takeuchi122,Schlegel11}, compressed 
sensing~\cite{Kudekar10,Krzakala12,Donoho12}, 
and models in statistical physics~\cite{Hassani12}.  
See \cite{Takeuchi122,Takeuchi121} for a theoretical treatment of general 
systems. 
 
A spatially coupled (SC) system is constructed as a one-dimensional chain of 
$L$ {\em large} subsystems. A slightly irregular structure, which results in 
a rate loss, is introduced at both ends of the chain so that information at 
both ends can be detected successfully. BP detection consists of two stages: 
training and propagation stages. In the training stage, information at both 
ends is first detected by utilizing the irregularity. We refer to positions 
at which irregularity is imposed as training positions. In the propagation 
stage, on the other hand, the reliable information at both ends propagates 
toward the center of the chain. {\em Order} preserved in each large subsystem 
enables no error propagation. Since the rate loss due to the irregularity 
at both ends vanishes as $L\to\infty$, the best performance 
is achieved\footnote{Reliable information fails to propagate with a finite 
probability when the size of each subsystem is finite. Consequently, an 
infinitely long chain may not be best for finite-sized systems.} in that 
limit. This implies that a long chain should be used to approach the MAP 
performance. 

When a long chain is used, many iterations are required for 
propagating reliable information over the whole chain. The purpose of this 
letter is to propose a novel method of coupling for accelerating the 
convergence of iterations. A remarkable effort oriented in the same direction 
was made by Truhachev et al.: They proposed to divide a long chain into several 
short chains and to re-connect the divided chains in a 
ladder-like~\cite{Truhachev121} or loop-like~\cite{Truhachev122} structure. 
The connecting positions, along with the end positions, correspond to the 
training positions. Reliable information at the connecting positions spreads 
over the chains simultaneously. As a result, the propagation of reliable 
information over the whole system is accelerated compared to one long chain. 

In this letter, we shall accelerate iterative detection in the training 
stage by creating {\em shortcuts} between distant training positions. Reliable 
information at each training position propagates to the other training  
positions through the shortcuts, whereas there is no collaboration between 
distant training positions for conventional SC systems. This collaborative 
training accelerates iterative detection in the training stage. 

This letter is organized as follows: After summarizing the notation and 
terminology used in this letter, in Section~\ref{sec2} we focus on SC 
CDMA systems as an example of SC systems, and explain how to connect 
training positions. Section~\ref{sec3} presents the density-evolution (DE) 
analysis of BP detection. In Section~\ref{sec4}, comparisons between 
collaborative training and non-collaborative training are made in terms of 
the number of iterations. Section~\ref{sec5} concludes this letter. 

\IEEEpubidadjcol

For integers~$l$ and $L$, $(l)_{L}$ is equal to $l+iL$ for an integer~$i$ 
such that $0\leq l+iL \leq L-1$. The set of consecutive integers 
$\{i,i+1,\ldots,j\}$ ($i<j$) is written as $[i:j]$. 
The vector $\boldsymbol{1}_{n}$ denotes 
the $n$-dimensional vector whose elements are all one. The Q-function 
$Q(\cdot)$ is the upper tail probability of the standard real Gaussian 
distribution. The real Gaussian distribution with mean~$\boldsymbol{m}$ and 
covariance~$\boldsymbol{\Sigma}$ is written as 
$\mathcal{N}(\boldsymbol{m},\boldsymbol{\Sigma})$. In a graph, the degree of 
a node is defined as the number of edges connected to the node. The distance 
between two nodes is the number of edges in the shortest path that connects 
the two nodes. 

\section{System Model} 
\label{sec2} 
\subsection{Spatially Coupled CDMA Systems} 
We consider synchronous $K$-user SC CDMA systems over the real additive white 
Gaussian noise (AWGN) channel as an example of SC systems. Note that it is 
possible to apply the idea in this letter to the other SC systems. For 
simplicity in presentation, {\em dense} spreading sequences are used, whereas 
the {\em dense} limit of sparse spreading sequences should be taken for 
rigorous DE~\cite{Montanari06,Guo08}. See \cite{Takeuchi122} for the details. 

Transmission over $L$ symbol periods is considered. Let $N_{l}$ and 
$\bar{N}=L^{-1}\sum_{l=0}^{L-1}N_{l}$ denote the spreading factor in symbol 
period~$l$ and the average spreading factor, respectively. The received vector 
$\boldsymbol{y}=(\boldsymbol{y}_{0}^{\mathrm{T}},\ldots,
\boldsymbol{y}_{L-1}^{\mathrm{T}})^{\mathrm{T}}\in\mathbb{R}^{L\bar{N}}$ is given by 
\begin{equation} \label{CDMA} 
\boldsymbol{y}
= \boldsymbol{S}\boldsymbol{x} + \boldsymbol{w}, 
\quad \boldsymbol{w}\sim\mathcal{N}(\boldsymbol{0},\sigma^{2}\boldsymbol{I}). 
\end{equation}
In (\ref{CDMA}), $\boldsymbol{x}=(\boldsymbol{x}_{0}^{\mathrm{T}},\ldots,
\boldsymbol{x}_{L-1}^{\mathrm{T}})^{\mathrm{T}}\in\{1,-1\}^{LK}$ denotes the 
data symbol vector with binary phase shift keying (BPSK), in which  
$\boldsymbol{x}_{m}\in\{1,-1\}^{K}$ consists of the $m$th data symbols of all 
users. Furthermore, $\boldsymbol{S}$ represents the $L\bar{N}\times LK$ 
spreading matrix, constructed as 
\begin{equation} \label{spreading_matrix} 
\boldsymbol{S} 
= \begin{bmatrix}
b_{0,0}\boldsymbol{S}_{0,0} & \cdots & b_{0,L-1}\boldsymbol{S}_{0,L-1} \\ 
\vdots & & \vdots \\ 
b_{L-1,0}\boldsymbol{S}_{L-1,0} & \cdots & b_{L-1,L-1}\boldsymbol{S}_{L-1,L-1} 
\end{bmatrix}. 
\end{equation}
In (\ref{spreading_matrix}), $\{\boldsymbol{S}_{l,m}\}$ are independent 
$N_{l}\times K$ matrices that have independent entries taking 
$\pm 1/\sqrt{N_{l}}$ with probability~$1/2$. Furthermore, 
$b_{l,m}\in\mathbb{R}$ is the $(l,m)$-element of the base matrix 
$\boldsymbol{B}\in\mathbb{R}^{L\times L}$, which characterizes spatial 
coupling. Imposing $\sum_{l=0}^{L-1}b_{l,m}^{2}=1$ for all $m$ normalizes the 
average power used for transmitting each data symbol.  

The set $\mathcal{L}=[0:L-1]$ of all symbol periods is decomposed 
into the training phase $\mathcal{T}\subset\mathcal{L}$ and the propagation 
phase $\mathcal{P}\subset\mathcal{L}$, which are disjoint subsets of 
$\mathcal{L}$. Let $N_{\mathrm{tr}}$ and $N$ denote the spreading factors 
in the training and propagation phases, respectively. Large $N_{\mathrm{tr}}$ 
is assumed to obtain reliable estimates of the data symbols transmitted 
in the training phase. The average load $\bar{\alpha}$ is defined as 
\begin{equation} \label{average_load} 
\bar{\alpha} = \frac{LK}{|\mathcal{T}|N_{\mathrm{tr}} + |\mathcal{P}|N} 
= \left\{
 \alpha_{\mathrm{tr}}^{-1}\frac{\tau}{L} 
 + \alpha^{-1}\left(
  1 - \frac{\tau}{L} 
 \right)
\right\}^{-1}, 
\end{equation}
with $\tau=|\mathcal{T}|$. In (\ref{average_load}), 
$\alpha_{\mathrm{tr}}=K/N_{\mathrm{tr}}$ and $\alpha=K/N$ denote the loads 
in the training and propagation phases, respectively. Under the BPSK 
assumption the average load~(\ref{average_load}) is equal to the average sum 
rate, and tends toward $\alpha$ as $L\to\infty$ and $\tau/L\to0$. 

\subsection{Spatial Coupling \& Collaborative Training} 
In one-dimensional or circular coupling with coupling width~$W$, a 
circulant matrix is used as the base matrix~$\boldsymbol{B}$: 
\begin{equation}
b_{l,m} = \frac{1}{\sqrt{2W+1}}b_{(l-m)_{L}},  
\end{equation}
with $(b_{0},\ldots,b_{L-1})=(\boldsymbol{1}_{W+1}^{\mathrm{T}},\boldsymbol{0},
\boldsymbol{1}_{W}^{\mathrm{T}})$. This regular base matrix is written as 
$\boldsymbol{B}_{L,W}^{(\mathrm{reg})}$. In order to present collaborative 
training, we shall introduce a graph representation of the base matrix 
$\boldsymbol{B}$. 

\begin{figure}[t]
\begin{center}
\includegraphics[width=\hsize]{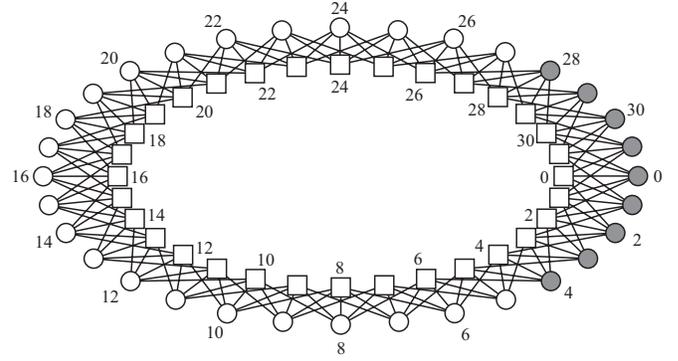}
\end{center}
\caption{
Regular bipartite graph with $L=32$ and $W=2$. 
}
\label{fig1} 
\end{figure}

The $L\times L$ base matrix~$\boldsymbol{B}$ can be represented by a 
bipartite graph that consists of $L$ factor nodes and $L$ variable nodes. 
The factor nodes shown by squares correspond to the row indices 
of $\boldsymbol{B}$, whereas the variable nodes represented by circles 
are associated with the column indices. If the element $b_{l,m}$ is non-zero, 
there is an edge between factor node~$l$ and variable node~$m$. Otherwise, 
there is no edge between the two nodes. In this letter, we refer to the 
bipartite graph that represents the regular base matrix 
$\boldsymbol{B}_{L,W}^{(\mathrm{reg})}$ as the regular bipartite 
graph $\boldsymbol{B}_{L,W}^{(\mathrm{reg})}$. See Fig.~\ref{fig1} for the 
regular bipartite graph~$\boldsymbol{B}_{32,2}^{(\mathrm{reg})}$. 

As an ensemble of collaborative training, we use an ensemble based on a 
small-world (SW) network~\cite{Watts98}, which is a model for elucidating the 
so-called SW phenomenon such as ``six degrees of separation.'' SW networks are 
highly clustered, as regular networks are, {\em and} have 
short path lengths between any two nodes, as random networks do. These 
properties of SW networks are suitable for accelerating iterative detection 
in the training stage. In this letter, this SW-network-based ensemble is 
referred to as ``SW ensemble.'' 

Let $\mathcal{C}_{W}^{(\mathrm{reg})}(m)\subset\mathcal{L}$ denote the set of 
node~$m$ and nodes with distance~$2$ from node~$m$ in a regular 
bipartite graph with coupling width~$W$. For example, 
$\mathcal{C}_{2}^{(\mathrm{reg})}(0)$ for variable nodes is shown by the gray 
nodes in Fig.~\ref{fig1}. The SW ensemble is obtained by modifying a regular 
bipartite graph at positions included in equally spaced $c$ clusters 
$\{\mathcal{C}_{W}^{(\mathrm{reg})}(iL/c)\}_{i=0}^{c-1}$. One could add nodes 
with more than two distances from node~$m$ into 
$\mathcal{C}_{W}^{(\mathrm{reg})}(m)$. However, such a generalization would 
increase the number of instances included in the SW ensemble. In this letter, 
only nodes with distance~$2$ from node~$m$ are included into 
$\mathcal{C}_{W}^{(\mathrm{reg})}(m)$. 
\begin{enumerate}
\item Let $i=0$ and generate the regular bipartite graph 
$\boldsymbol{B}_{L,W}^{(\mathrm{reg})}$. 
\item Repeat the following for all variable nodes~$m\in
\mathcal{C}_{W}^{(\mathrm{reg})}(iL/c)$ in the $i$th cluster: 
With probability~$p$, re-connect each edge that is connected to variable 
node~$m$ to a factor 
node~$l\in\cup_{j\neq i}\mathcal{C}_{W}^{(\mathrm{reg})}(jL/c)$ in the 
other clusters uniformly and randomly. 
\item If $i=c-1$, terminate the algorithm. Otherwise, go back to Step~2) 
after $i:=i+1$. 
\end{enumerate} 

The probability~$p$ controls a tradeoff between the {\em clustering} property 
and the short-path-length property: The SW coupling with $p=0$ reduces to 
the highly-clustered regular coupling, whereas it with $p=1$ does to the 
random coupling with short path lengths. Thus, moderate $p$ should be used 
to obtain instances with the two properties. 

\begin{figure}[t]
\begin{center}
\includegraphics[width=\hsize]{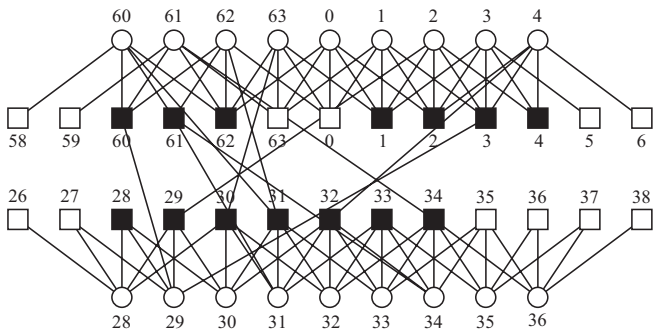}
\end{center}
\caption{
A bipartite graph obtained from the $(64,2,0.1,2,14)$-SW ensemble. The black 
factor nodes show the nodes corresponding to the training phases. 
The variable nodes $m\notin C_{2}^{(\mathrm{reg})}(0)\cup 
C_{2}^{(\mathrm{reg})}(32)$ and the associated factor nodes are omitted, since 
they can be restored uniquely. 
}
\label{fig2} 
\end{figure}

In bipartite graphs obtained by the algorithm above, different factor nodes 
may have different degrees, whereas all variable nodes have degree~$2W+1$.  
We propose a heuristic algorithm for determining the training 
phase $\mathcal{T}$ with size $\tau=|\mathcal{T}|$. The proposed algorithm 
assigns factor nodes with largest degrees to the training phase. 
Factor nodes with large degrees serve as hubs from which reliable information 
spreads over variable nodes. 
\begin{enumerate}
\item Let $\tilde{\tau}=\tau$, $\mathcal{T}=\emptyset$, and $d=d_{\mathrm{max}}$, 
with $d_{\mathrm{max}}$ denoting the maximum degree of the factor nodes. 
\item If the number of factor nodes with degree~$d$ is greater than or equal 
to $\tilde{\tau}$, pick up $\tilde{\tau}$ factor nodes from 
all factor nodes with degree~$d$ uniformly and randomly, add the 
corresponding indices into $\mathcal{T}$, and terminate the algorithm.  
Otherwise, go to the next step. 
\item Add the indices $\mathcal{L}_{d}$ that represent all factor nodes with 
degree~$d$ into $\mathcal{T}$, and go back to Step~2) after 
$\tilde{\tau}:=\tilde{\tau}-|\mathcal{L}_{d}|$ and $d:=d-1$. 
\end{enumerate} 
We refer to the ensemble\footnote{
The ensemble with $L=64$, $W=2$, and $c=2$ contains all bipartite graphs 
obtained by re-connection of the edges in Fig.~\ref{fig2}. However, their 
occurrence probabilities are non-uniform and determined on the basis of the 
probability~$p$, whereas instances contained in conventional ensembles are 
assumed to occur uniformly and randomly.} that consists of all instances 
obtained by the two algorithms above as the $(L,W,p,c,\tau)$-SW ensemble. 
See Fig.~\ref{fig2} for an instance obtained from the SW ensemble. 

\begin{remark}
The degrees of the factor nodes provide an impact on the complexity of 
iterative detection algorithms. One method for making a fair comparison 
between regular and irregular bipartite graphs would be to consider sparsely 
spread CDMA systems in which, for factor nodes with a large degree, small row 
weights are assigned to the corresponding subsystems. However, this influence 
on the performance vanishes in the dense limit. This argument implies that 
comparisons presented in this letter are not necessarily unfair in terms 
of the complexity. 
\end{remark} 

\section{Density Evolution} 
\label{sec3} 
We consider an iterative detection algorithm based on BP with the Gaussian 
approximation~\cite{Kabashima03}. See \cite{Takeuchi122} for the details. 
In order to evaluate the bit error rate (BER) via DE, we take 
the large-system limit in which $K$ and $\{N_{l}\}$ tend to infinity with 
the ratios $\alpha_{l}=K/N_{l}$ fixed for all $l$. The following theorem 
is useful for quickly evaluating the performance of instances picked up from 
the SW ensemble. 

\begin{theorem} \label{theorem1} 
In the large-system limit, the BER of each element of the $m$th data symbol 
vector $\boldsymbol{x}_{m}$ in iteration~$i$ is given by 
$Q(\sqrt{\mathrm{sir}_{m}(i)})$, in which $\{\mathrm{sir}_{m}(i)\}$ are 
determined via the coupled DE equations,  
\begin{equation} \label{DE1} 
\mathrm{sir}_{m}(i) = \sum_{l=0}^{L-1}\frac{b_{l,m}^{2}}{\sigma_{l}^{2}(i)},  
\end{equation}
\begin{equation} \label{DE2} 
\sigma_{l}^{2}(i) = \sigma^{2} + \alpha_{l}\sum_{m=0}^{L-1}b_{l,m}^{2} 
\mathrm{MMSE}(\mathrm{sir}_{m}(i-1)), 
\end{equation}
with $\mathrm{sir}_{m}(0)=0$ for all $m$. In (\ref{DE2}), the function 
$\mathrm{MMSE}(x)$ denotes the minimum mean-squared error (MMSE) of the 
BPSK-input AWGN channel with signal-to-noise ratio (SNR)~$x$. 
\end{theorem} 
\begin{IEEEproof}
The proof is a generalization of the results in \cite{Takeuchi122} and 
is therefore omitted. 
\end{IEEEproof} 

When a bipartite graph of coupling and the load $\alpha_{\mathrm{tr}}$ 
in the training phase are given, the DE equations~(\ref{DE1}) and 
(\ref{DE2}) have a unique fixed-point for small SNR~$1/\sigma^{2}$ or 
small load $\alpha$ in the propagation phase. On the other hand, 
there may be multiple fixed-points for large SNR and large load. 
The BP threshold for fixed SNR is defined as the maximum load such that 
the DE equations have a unique fixed-point. 
\begin{definition}
When $\alpha_{\mathrm{tr}}$, a bipartite graph of coupling, and SNR are 
fixed, the BP threshold is defined as the supremum of $\alpha_{\mathrm{c}}$ 
such that the DE equations~(\ref{DE1}) and (\ref{DE2}) have a unique 
fixed-point for all $\alpha<\alpha_{\mathrm{c}}$. 
\end{definition} 
See \cite{Takeuchi122} for how to estimate the BP threshold. 
The operational meaning of the BP threshold is that, when $\alpha$ is smaller 
than the BP threshold, the BP-based iterative algorithm can eliminate 
multiple-access interference (MAI), and achieve performance close to the 
single-user bound. On the other hand, the system is MAI-limited when 
$\alpha$ is larger than the BP threshold. When $1/\sigma^{2}=10$~dB, 
the BP threshold $\alpha_{\mathrm{BP}}$ for the uncoupled CDMA system is given 
by $\alpha_{\mathrm{BP}}\approx 1.73078$~\cite{Takeuchi122}. 
As $L$ and $W$ tend to infinity with $W/L\to0$, on the other hand, 
the corresponding BP threshold $\alpha_{\mathrm{BP}}^{(\mathrm{reg})}$ for 
the circularly coupled CDMA system tends toward the optimal threshold 
$\alpha_{\mathrm{MAP}}\approx1.98267$~\cite{Takeuchi122}. 

\section{Numerical Comparisons}
\label{sec4} 
We shall make comparisons between regular coupling and the SW ensemble. 
Bipartite graphs obtained from the SW ensemble have distinctly different 
performance instance by instance for small-sized graphs. Thus, we used 
good instances that are obtained by examining many instances picked up  
from the SW ensemble with Theorem~\ref{theorem1}. 

\begin{figure}[t]
\begin{center}
\includegraphics[width=\hsize]{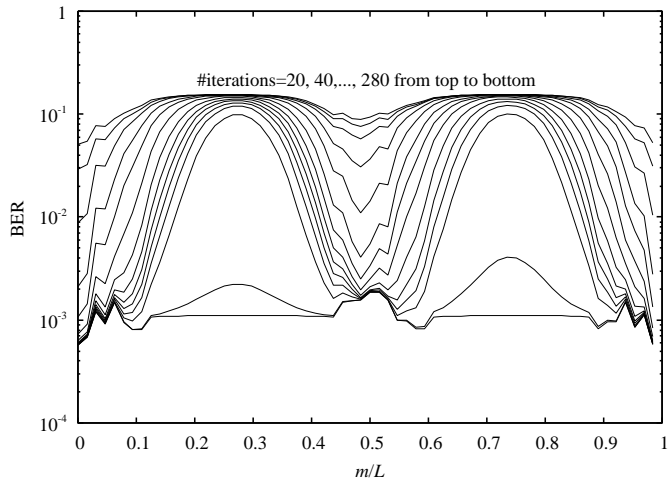}
\end{center}
\caption{
BER versus $m/L$ for the bipartite graph of coupling obtained from 
the $(64,2,0.1,2,14)$-SW ensemble, shown in Fig.~\ref{fig2}. 
$1/\sigma^{2}=10$~dB, $\alpha_{\mathrm{tr}}=1.45$, and $\alpha=1.98$. 
}
\label{fig3} 
\end{figure}

Figure~\ref{fig3} shows the evolution of the BERs for the bipartite graph of 
coupling shown in Fig.~\ref{fig2}. We find that the data symbols in the 
two clusters $\mathcal{C}_{2}^{(\mathrm{reg})}(0)$ and 
$\mathcal{C}_{2}^{(\mathrm{reg})}(32)$ are detected first. Their BERs are not 
constant, because of the irregularity of the bipartite graph in the clusters. 
Then, the reliable information propagates toward the middle positions 
between the two clusters. Eventually, the BERs at all positions 
tend to a small level of $10^{-3}$.  


\begin{figure}[t]
\begin{center}
\includegraphics[width=\hsize]{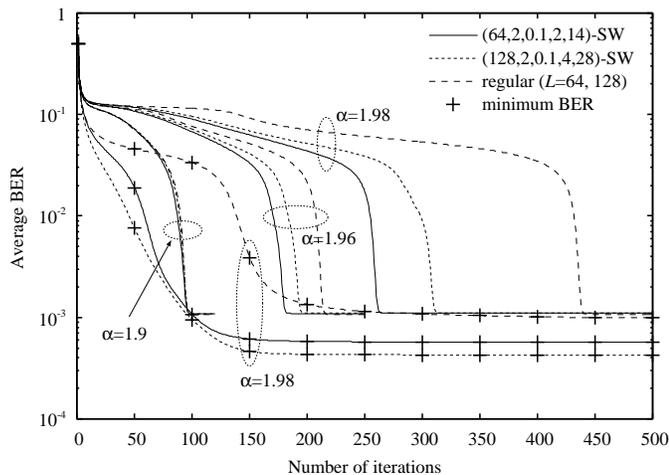}
\end{center}
\caption{
Average BER versus the number of iterations for $W=2$, $1/\sigma^{2}=10$~dB, 
and $\alpha_{\mathrm{tr}}=1.45$. For the regular coupling with $L=64$ 
(resp.\ $L=128$), 14 (resp.\ 28) symbol periods 
$\mathcal{T}=\{[61:63],[0:3],[29:35]\}$ (resp.\ 
$\mathcal{T}=\{[125:127],[0:3],[29:35],[61:67],[93:99]\}$) are assigned to the 
training phase. The bipartite graph of coupling shown in Fig.~\ref{fig2} was 
used for the $(64,2,0.1,2,14)$-SW ensemble. 
}
\label{fig4} 
\end{figure}

Figure~\ref{fig4} presents comparisons between the regular coupling and 
the SW ensembles. For $\alpha=1.9$ much smaller than the optimal threshold 
$\alpha_{\mathrm{MAP}}\approx1.98267$, three methods of coupling are 
indistinguishable from each other. For $\alpha=1.98$ close to the optimal 
threshold, on the other hand, the average BERs for the SW ensembles converge 
toward a small BER more quickly than that for the regular coupling. 
It is worth noting that the $(64,2,0.1,2,14)$-SW ensemble outperforms the 
$(128,2,0.1,4,28)$-SW ensemble. This implies that there is no point in using 
the SW ensembles with $c>2$, or that it is difficult to find a good instance 
from such SW ensembles. In order to investigate the reasons of the quick 
convergence, we focus on the minimum BERs over all $m=0,\ldots,L-1$, shown 
by the lines with pluses, which should be at a training position. 
We find that the minimum BERs for the SW ensembles converge more quickly 
than that for the regular coupling. This quick convergence in the training 
phase reduces the overall number of iterations. Another reason is a slight 
improvement of the BP threshold: The BP thresholds for the regular coupling 
and the $(64,2,0.1,2,14)$-SW ensemble are approximately  
$1.98958$ ($\bar{\alpha}\approx1.83981$) and 
$1.99911$ ($\bar{\alpha}\approx1.84617$), respectively.  

\section{Conclusions} 
\label{sec5} 
The SW ensemble of coupling has been proposed to accelerate iterative 
detection in the training stage for SC CDMA systems. Instances picked up 
from the ensemble have direct connections between distant training positions.   
The DE analysis has shown that the proposed method can provide a significant 
reduction in the number of iterations for highly loaded systems compared 
to the conventional method of coupling. We conclude that irregular coupling 
in the training phase can accelerate iterative detection for SC systems. 



\ifCLASSOPTIONcaptionsoff
  \newpage
\fi



\bibliographystyle{IEEEtran}
\bibliography{IEEEabrv,kt-coml2012}
\end{document}